\documentclass[epj]{webofc}
\usepackage[varg]{txfonts}   % Web of Conferences font
\usepackage{graphicx}
\usepackage{epstopdf}
\usepackage{amsmath,amssymb}

\newcommand{\beq}{\begin{equation}}
\newcommand{\eeq}{\end{equation}}
\newcommand{\bea}{\begin{eqnarray}}
\newcommand{\eea}{\end{eqnarray}}
\newcommand{\nn}{\nonumber}

\newcommand{\tr}{{\rm Tr}}

\def\lsi{\raise0.3ex\hbox{$<$\kern-0.75em\raise-1.1ex\hbox{$\sim$}}}
\def\gsi{\raise0.3ex\hbox{$>$\kern-0.75em\raise-1.1ex\hbox{$\sim$}}}
\newcommand{\losim}{\mathop{\lsi}}

\wocname{EPJ Web of Conferences}
\woctitle{CONF12}
\begin{document}
\selectlanguage{english}
\title{Towards a theoretical description of dense QCD}
%
% subtitle (optional, strongly discouraged)

\author{%Jonas Glesaaen\inst{1,2}\fnsep\thanks{\email{Mail address for first author}} \and
        Owe Philipsen\inst{1}\fnsep\thanks{\email{philipsen@th.physik.uni-frankfurt.de}}         % etc.
}

\institute{%Swansea
%\and
           Institut f\"ur Theoretische Physik, Goethe-Universit\"at Frankfurt, 60438 Frankfurt, Germany 
}

\abstract{%
The properties of matter at finite baryon densities play an important role for the astrophysics of compact stars as well as for heavy 
ion collisions or the description of nuclear matter. Because of the sign problem of the quark determinant, lattice QCD cannot be 
simulated by standard Monte Carlo at finite baryon densities. I review alternative attempts to treat dense QCD with an effective  
lattice theory derived by analytic strong coupling and hopping expansions, 
which close to the continuum is valid for heavy quarks only, but shows all qualitative features of nuclear physics emerging from QCD. 
In particular, the nuclear liquid gas transition and an equation of state for baryons can be calculated directly from QCD. 
A second effective theory based on strong coupling methods permits studies of the phase diagram in the chiral limit on 
coarse lattices.
}
\maketitle
\section{Introduction}
\label{intro}
The understanding of the different forms of nuclear matter under extreme conditions plays an 
increasingly important role for nuclear astrophysics, particle physics and heavy ion collisions. 
Unfortunately, the infamous "sign problem" of QCD, i.e.~the fact that the fermion determinant
becomes complex with real chemical potential for baryon number $\mu_B$, 
prohibits direct Monte Carlo simulations of lattice QCD at finite matter density. 
Approximate methods are able to circumvent this problem
only for small quark chemical potentials $\mu=\mu_B/3\losim T$ \cite{review}.
So far, no sign of a critical point or a first order phase transition has been found in this controlled 
region. Complex Langevin algorithms do not suffer 
from the sign problem, but occasionally converge to incorrect answers and thus have their 
own problems. A lot of progress has been made recently, but no non-analytic phase transition has been 
reported in this approach so far \cite{langevin}.

These difficulties motivate the development of effective theories whose sign problem is mild enough to
simulate the cold and dense region of QCD. In this contribution I summarise two approaches based on 
analytic strong coupling expansions of lattice QCD, that achieve this goal in two complementary 
parameter regions, either for QCD with very heavy quarks in the continuum,
or for QCD with light quarks on coarse lattices.
The heavy dense region is also studied by complex Langevin simulations \cite{sexty} and the density of states 
method \cite{lang}, for
which the effective theory results can serve as a benchmark. Furthermore, the effective
theory approach can be tested against full simulations in two-colour QCD \cite{scior2}.

\section{A 3d effective lattice theory for heavy quarks}
\label{sec-1}

\begin{figure}
\centering
\includegraphics[width=0.9\textwidth]{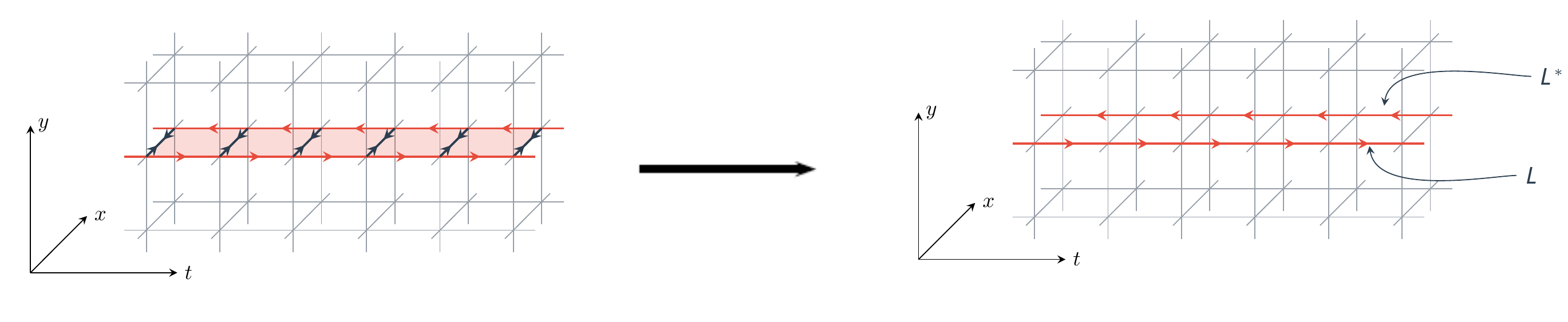}
\caption[]{Integrating over spatial links transforms the slice of lattice action into a nearest neighbour 
interaction of a Polyakov loop and the neighbouring conjugate Polyakov loop.}
\label{fig:zn}
\end{figure}
Starting point is the path integral over 
Wilson's formulation of lattice QCD at finite temperature $T=(aN_\tau)^{-1}$. Integrating over the spatial lattice link variables 
then defines an effective theory as a function of temporal link variables only. Because of (anti-)periodic boundary conditions
for (fermions) bosons, the resulting effective theory can be formulated as a theory of Polyakov loops 
$L({\bf x})=\tr\prod_{\tau=1}^{N_\tau} U_0({\bf x},\tau)$,
\beq
Z=\int DU_0DU_i\;det Q \; e^{-S_g[U]}\equiv\int DU_0\;e^{-S_{eff}[U_0]}=\int DL \,\;e^{-S_{eff}[L]}\;.
\eeq 
Without truncations, the effective action is unique and exact. 
Since all spatial links are
integrated over, the resulting effective action has long-range interactions 
of Polyakov loops at all distances and to all powers so that in practice truncations are necessary. 
For non-perturbative ways to define and determine truncated 
theories, see \cite{wozar,green1,green2,bergner}.
Here we expand the path integral in a combined strong coupling and
hopping expansion, with interaction
terms ordered according to their leading powers in the lattice gauge coupling $\beta$ and the hopping parameter $\kappa$,
\beq
\beta=\frac{2N_c}{g^2}, \quad \kappa=\frac{1}{2am_q+8}\;.
\eeq
The hopping expansion is effectively in inverse quark mass and this is the reason that the effective theory is valid for heavy quarks only.
For the gauge action it is advantageous to perform a character expansion in terms of the fundamental character
$u(\beta)=\beta/18+\beta^2/216+O(\beta^3)< 1$.
Its dependence on the lattice gauge coupling $\beta$ is known to arbitrary precision and
it is always smaller than one for finite $\beta$-values. Both expansions result in convergent series within a finite convergence
radius.
Truncating these at some finite order, integration over the
spatial gauge links can be performed analytically to provide a closed expression for the effective theory.
To leading order the effective gauge action reads
\begin{eqnarray}
e^{S_{\mathrm{eff}}^{(1)}}=\lambda(u,N_\tau)\sum_{<ij>}\left(L_iL_j^\ast+L_i^\ast
L_j\right)\;,
\qquad\lambda(u,N_\tau)=u^{N_\tau}\Big[1+\ldots\Big]\;,
%\lambda(u,N_{\tau}\geq5)&=&u^{N_\tau}\exp\bigg[N_{\tau}\bigg(4u^4+12u^5-14u^6-36u^7
%		+\frac{295}{2}u^8+\frac{1851}{10}u^9+\frac{1055797}{5120}u^{10}+\ldots\bigg)\bigg]\;,
\label{eq_lambda}
\end{eqnarray}
where higher order corrections to $\lambda(u,N_\tau)$ and 
higher order 
interaction terms can be found in \cite{efft1}. Going via an effective action
results in a resummation to all powers with better convergence properties compared to a direct series expansion of 
thermodynamic observables as in \cite{lange1,lange2}. 
Since the Polyakov loop $L({\bf x})$ contains the length $N_\tau$ of the 
temporal lattice extent implicitly, the effective theory is three-dimensional.
It is arranged as
\begin{eqnarray}
-S_{\mathrm{eff}}=\sum_{i=1}^\infty\lambda_i(u,\kappa,N_\tau)S_i^s-
2N_f\sum_{i=1}^\infty\left[h_i(u,\kappa,\mu,N_\tau)S_i^a+
\bar{h}_i(u,\kappa,\mu,N_\tau)S_i^{a,\dagger}\right]\;.
\label{eq_defseff}
\end{eqnarray}
The $\lambda_i$ are defined as the effective couplings of the 
$Z(3)$-symmetric terms $S_i^s$, whereas the $h_i$ multiply the asymmetric 
terms $S_i^a$.
The $h_i$ and $\bar{h}_i$ are related via
$\bar{h}_i(u,\kappa,\mu, N_\tau)=h_i(u,\kappa,-\mu, N_\tau)$.

\subsection{Testing the effective theory at zero density}

The effective theory can be tested against the full one in a few cases at zero density.
First we investigate the deconfinement transition of Yang-Mills theory, i.e.~all the centre symmetry breaking couplings $h_i=\bar{h_i}=0$.
The simplest effective theory consists of just one nearest neighbour coupling between Polyakov loops, which has been 
computed through order $u^{10}$ and includes all higher powers by resumming some classes of higher order diagrams.
This simple effective theory exhibits a Z(3) centre symmetry breaking phase transition, associated with a rise of 
the Polyakov loop as shown in figure \ref{fig:zn} (left).  In the infinite volume limit this becomes a discontinuous jump 
indicating a first-order transition. The same effective theory evaluated for SU(2) shows indeed a second order
transition \cite{efft1}.
The critical coupling $\lambda_1^c$ 
can be translated to a lattice gauge coupling $\beta^c$
by inverting Equation (\ref{eq_lambda}) for every given $N_\tau$. Using a non-perturbative beta-function from the 
literature \cite{sommer}, the couplings are converted to critical temperatures and shown in figure \ref{fig:zn} (right). The error
bars are systematic and reflect the difference between the two highest orders of the effective action. Effects of subleading
couplings have been found to be small.
Note, that the predictions for $T_c(N_\tau)$ all follow by analytic mapping from one single critical effective coupling
$\lambda^c_1(u,N_\tau)$, 
in contrast to the full theory, where every $N_\tau$ has to be simulated. The scaling region with 
leading order $a^2\sim N_\tau^{-2}$ cut-off effects can be reached (corresponding to $\beta\sim 5-6.5$) and a continuum extrapolation is within 10\% of the 
full answer.
\begin{figure}
\centering
\includegraphics[width=7cm]{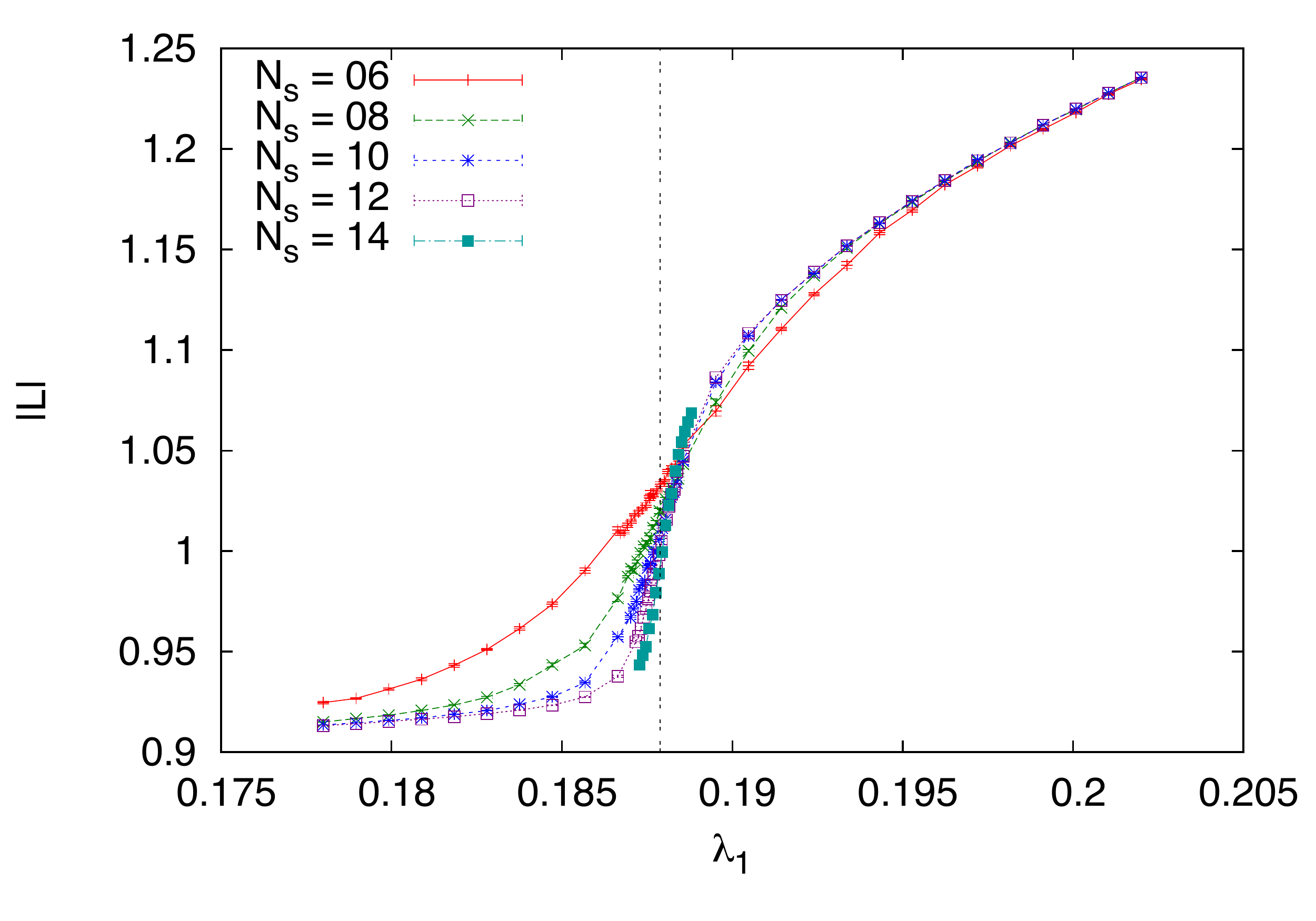}
\includegraphics[width=7cm]{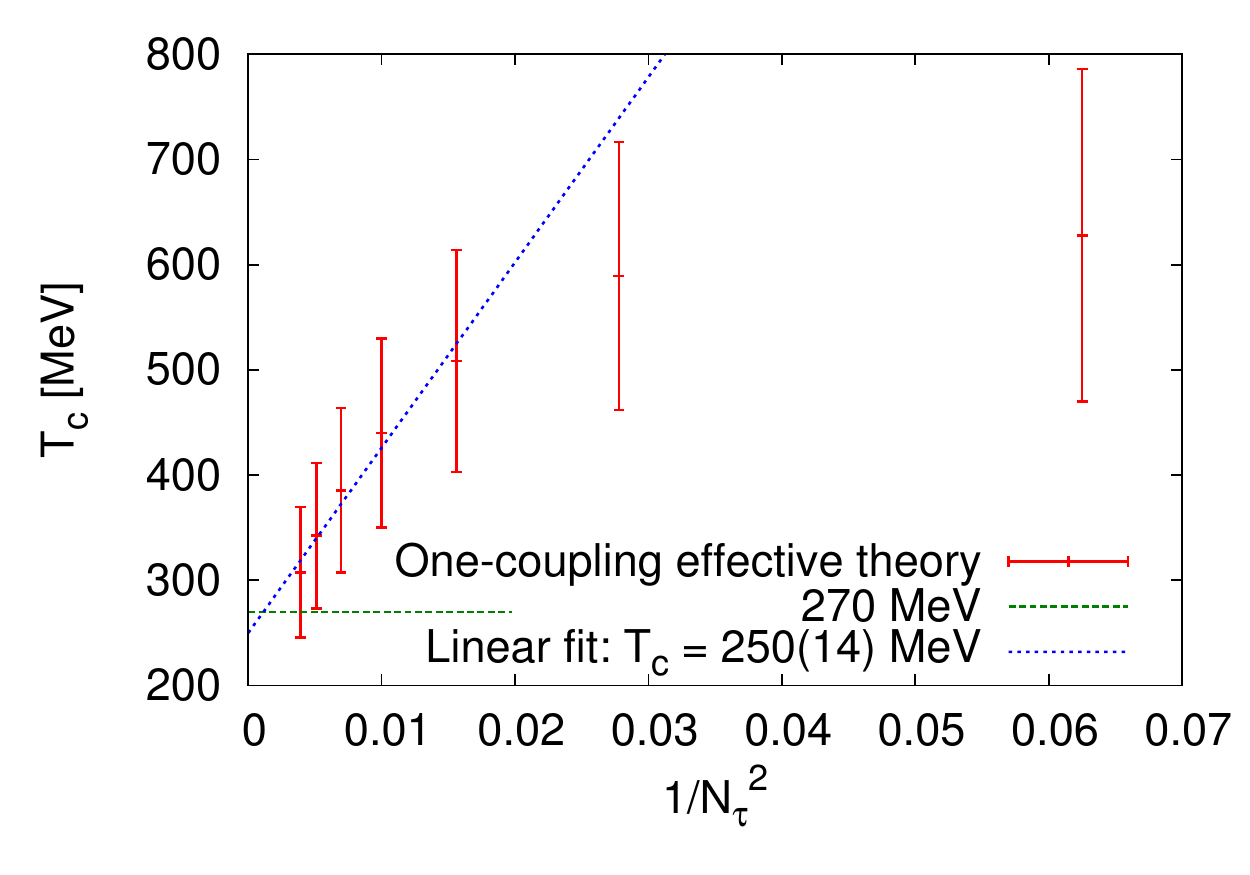}
\caption[]{Left: Expectation
value of $|L|$. The vertical line marks the infinite-volume transition. From \cite{efft1}. Right: Continuum approach of $T_c$ predicted
by the 3d effective theory. From \cite{kappa}.}
\label{fig:zn}
\end{figure}

Next consider QCD with dynamical but heavy quarks. These break the centre symmetry explicitly and the first order
transition weakens with decreasing quark mass until it is lost at some critical value. For still lighter quarks the deconfinement
transition is an analytic crossover. This behaviour is inherited by the effective theory and again reproduced faithfully, as
the comparison of critical hopping parameters between the 3d effective theory and 4d QCD in Table \ref{tab} shows.
\begin{table}[h]
\centering
\caption{Location of the critical point for $\mu=0$ and $N_\tau=4$. The first two columns are from the 3d effective theory, 
the last compares with existing literature.}
\label{tab}
\begin{tabular}{|c|c|c||c|}
\hline
	 $N_f$ & $M_c/T$ & $\kappa_c(N_\tau=4)$ & $\kappa_c(4)$, Ref.~\cite{saito}  \\
\hline
	1 & 7.22(5)  & 0.0822(11)           & 0.0783(4)  \\
	2 & 7.91(5)  & 0.0691(\phantom{0}9) & 0.0685(3)  \\
	3 & 8.32(5)  & 0.0625(\phantom{0}9) & 0.0595(3)   \\
\hline
\end{tabular}
\end{table}

\subsection{The deconfinement transition at finite density}

\begin{figure}[t]
\centering
\includegraphics[width=7cm,clip]{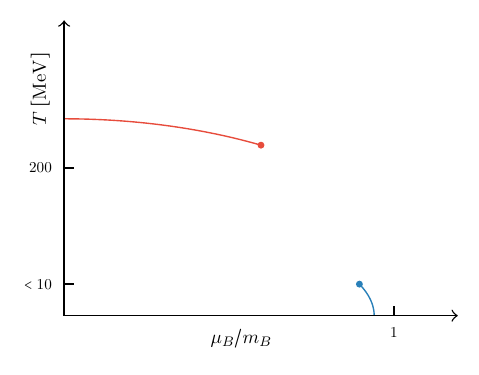}
\includegraphics[width=7cm,clip]{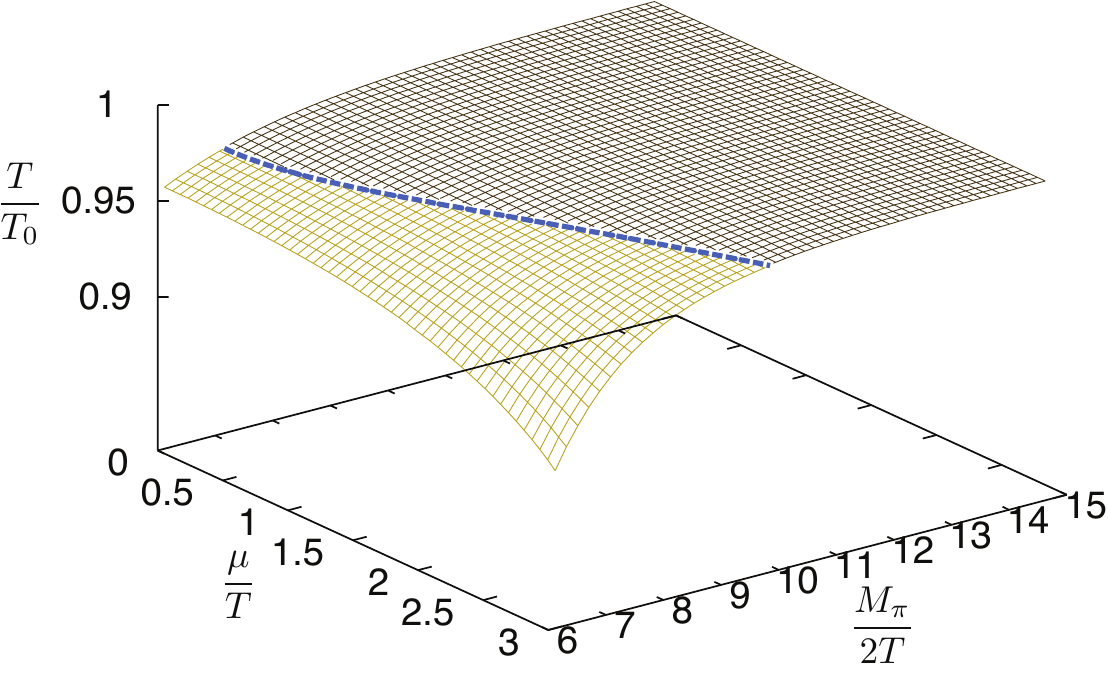}
\caption{Left: Qualitative phase diagram for QCD with very heavy quarks. Right: Phase diagram for $N_f=2, N_\tau=6$ 
from the 3d effective theory \cite{kappa}.}
\label{fig:heavy}       % Give a unique label
\end{figure}

After these successful comparisons with the full 4d theory, we can switch on a finite chemical potential and study how
the finite temperature deconfinement transition changes. The effective theory can be simulated by ordinary Monte Carlo 
and reweighting as well as complex Langevin without runaway solutions \cite{kappa,bind}. The resulting phase diagram is shown 
in figure \ref{fig:heavy}. The first order deconfinement transition is weakened
by a real chemical potential and disappears at a critical end point, which has been calculated as a function of pion mass \cite{kappa}.
This behaviour of the phase transition is also reproduced by continuum studies using a Polyakov loop model \cite{lo} and in the  
functional renormalisation group approach \cite{frg}.

\subsection{The nuclear liquid gas transition}

The most difficult region to address is that of cold and dense QCD, since the sign problem grows exponentially with $\mu/T$.  
It is instructive to consider the static and strong coupling limits. In this case the partition function 
factorises into one-site integrals, corresponding to a gas of free hadrons, that can be solved analytically. In the zero temperature limit for
$N_f=1$ \cite{silver, bind} we have
\beq
Z(\beta=0,\kappa=0) \stackrel{T\rightarrow 0}{\longrightarrow}\;
\left[1+4h_1^{N_c}+h_1^{2N_c}\right]^{N_s^3}\quad \mbox{with} \quad h_1=(2\kappa e^{a\mu})^{N_\tau}=e^{(m-\mu)/T}\;.
\eeq
The quark number density is now easily evaluated
\beq
n=
\frac{T}{V}\frac{\partial}{\partial \mu}\ln Z=\frac{1}{a^3}\frac{4N_ch_1^{N_c}+2N_ch_1^{2N_c}}{1+4h_1^{N_c}+h_1^{2N_c}}\;,
 \quad \lim_{T\rightarrow 0} a^3n=\left\{\begin{array}{cc} 0, & \mu<m\\
	2N_c, & \mu>m\end{array}\right.\;,
\eeq
and at zero temperature exhibits a discontinuity when the quark chemical potential equals the constituent mass $m$.
This reflects the ``silver blaze'' property of QCD, i.e.~the fact that the baryon number stays zero
for small $\mu$ even though the partition function explicitly depends on it \cite{cohen}. Once the baryon chemical potential 
$\mu_B=3 \mu$
is large enough to make a baryon ($m_B=3m$ in the static strong coupling limit), a transition 
to the lattice saturation density takes place. 
Note that saturation density here is $2N_c$ quarks per flavour and lattice
site and reflects the Pauli priniciple. This is clearly a discretization effect that has to disappear
in the continuum limit.

\begin{figure}[t]
\centering
\includegraphics[width=7cm]{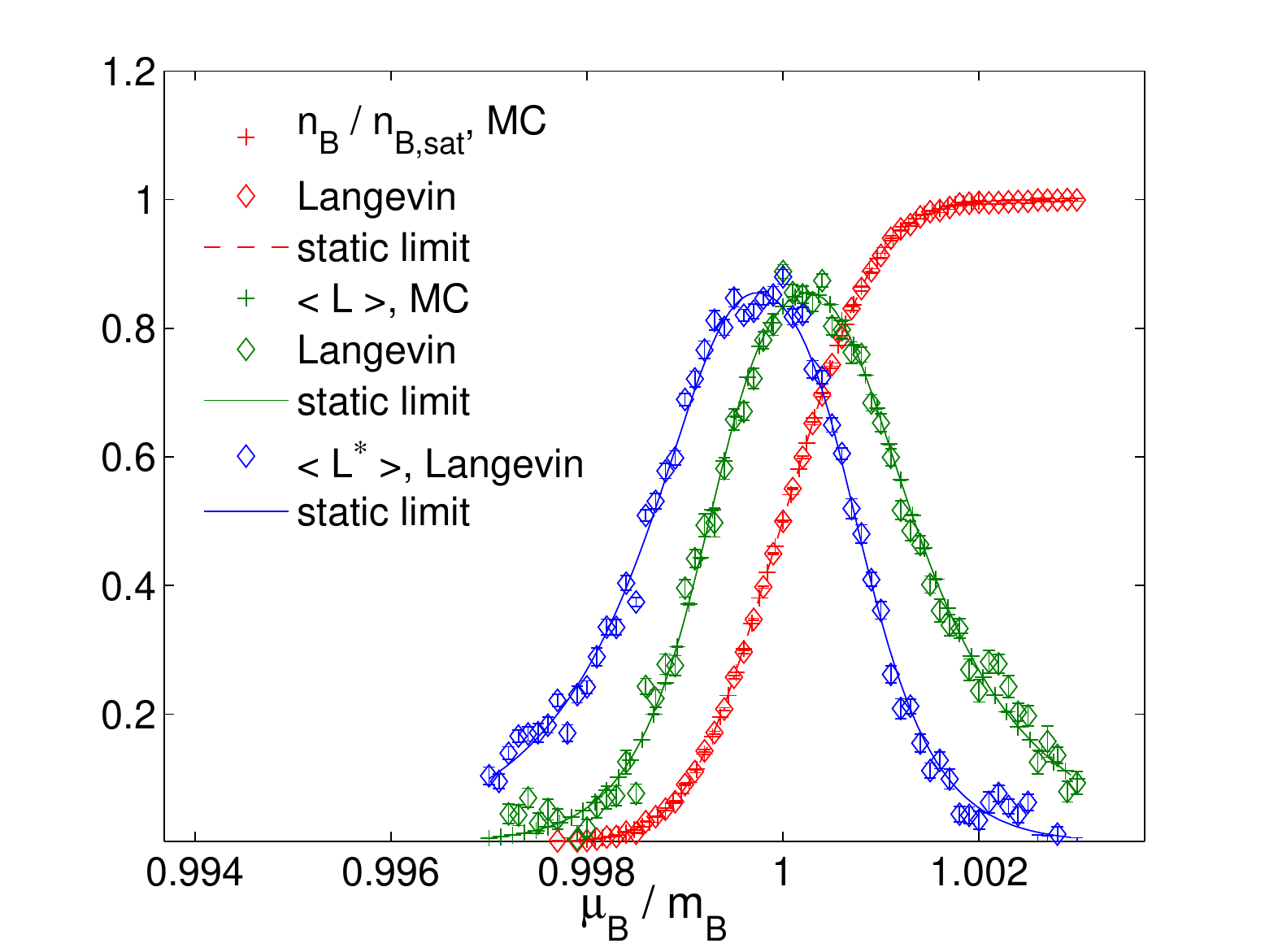}
\includegraphics[width=7cm]{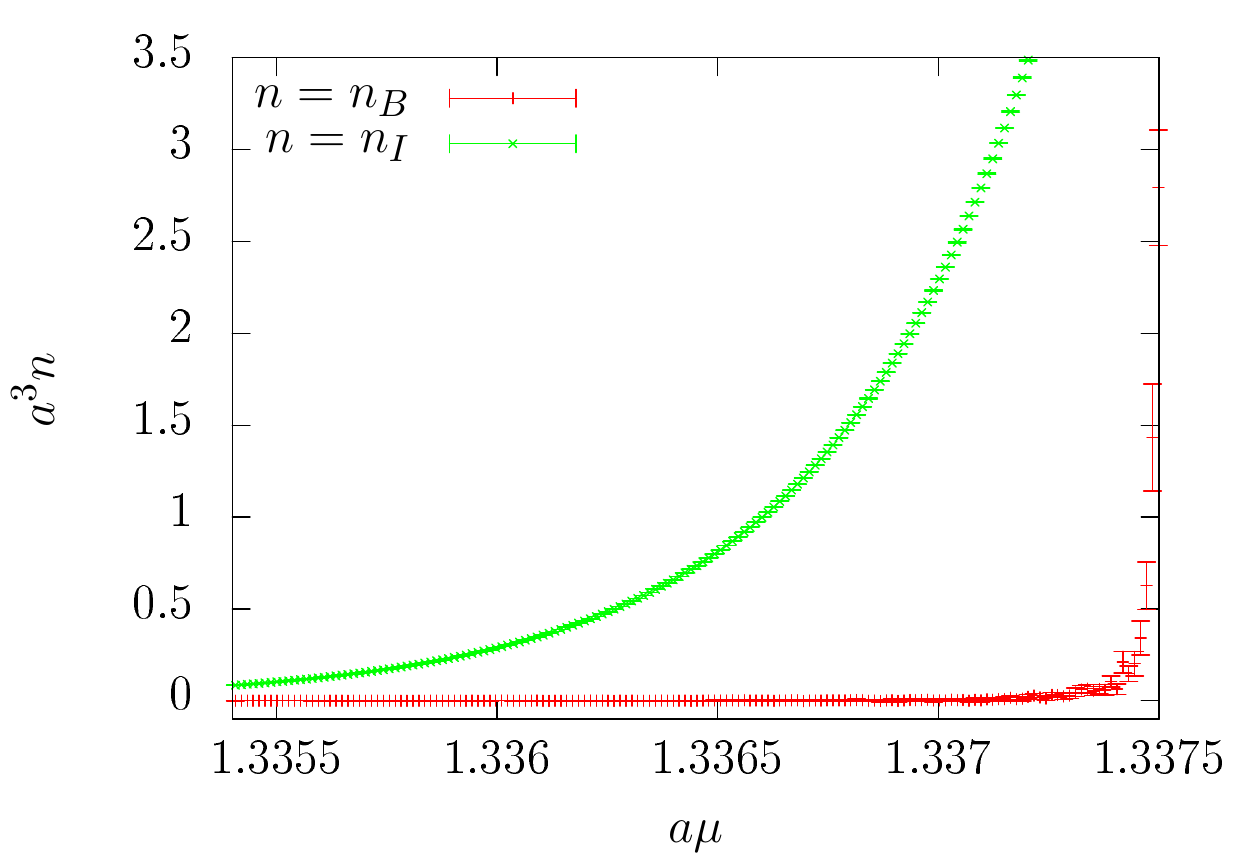}
\caption[]{Left: Baryon density, Polyakov loop and conjugate Polyakov loop 
obtained from Monte Carlo, complex Langevin and the static strong coupling limit, respectively. 
Right: Onset to isospin condensation and baryon condensation at finite chemical potential for isospin (green)
and bayon number (red), respectively. From \cite{bind}.}
\label{fig:dense}
\end{figure}

Next, we switch on the gauge coupling as well as the fermionic couplings $h_1,h_2$ through order
$\kappa^2$, where the latter also includes $L_iL_j$ terms and hence 
quark-quark-interaction (for explicit expressions, see \cite{silver}).
To keep our truncated series in full control, we choose 
$\beta=5.7, \kappa = 0.0000887, N_\tau=116$ corresponding to 
$m_M=20$ GeV, $T=10$ MeV, $a=0.17$ fm. The silver blaze property as well as lattice
saturation are observed also in the interacting case, but the step function is now smoothed, 
Figure \ref{fig:dense} (left).
Note that the Polyakov loop as well as its conjugate get screened in the presence of a 
baryonic medium, and hence rise. 
The ensuing decrease is due to lattice saturation which forces all $Z(3)$ states to be occupied. 
Note that this decrease happens before saturation near the
point of half filling. There is an approximate particle anti-particle symmetry about this point
and one expects artefacts to be dominant from that density upwards \cite{rind}. 
Figure \ref{fig:dense} (right) compares the onset transition at finite isospin chemical potential to that at
finite bayon chemical potential. In the static limit $m\rightarrow\infty$ both happen at the same quark chemical potential 
$\mu^c= m_B/3=m_\pi/2$. As the quark mass is lowered, a gap opens up between isospin
and baryon condensation in accord with the gap in the lowest states of the meson and baryon spectrum \cite{cohen}.

\begin{figure}[t]
\centering
\includegraphics[width=7cm]{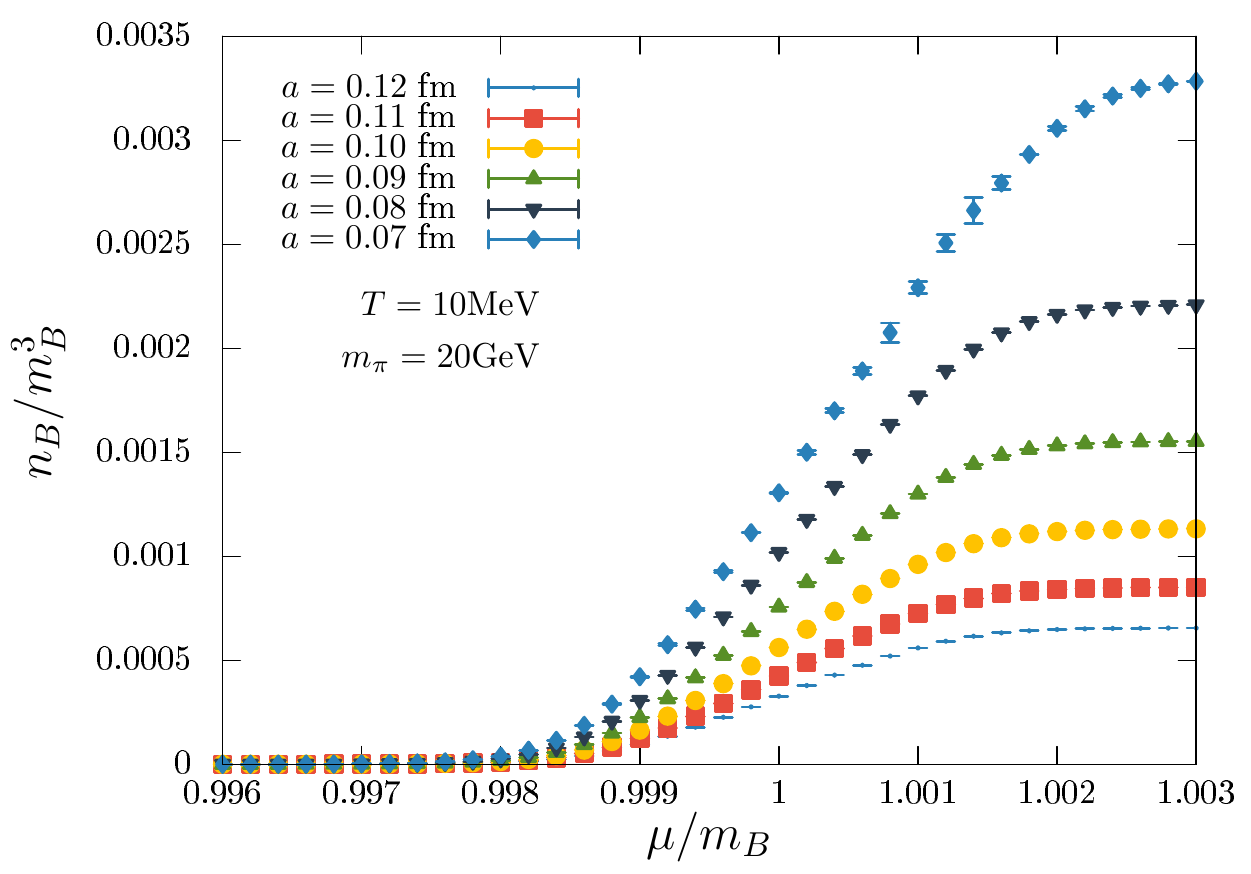}
\includegraphics[width=7cm,clip]{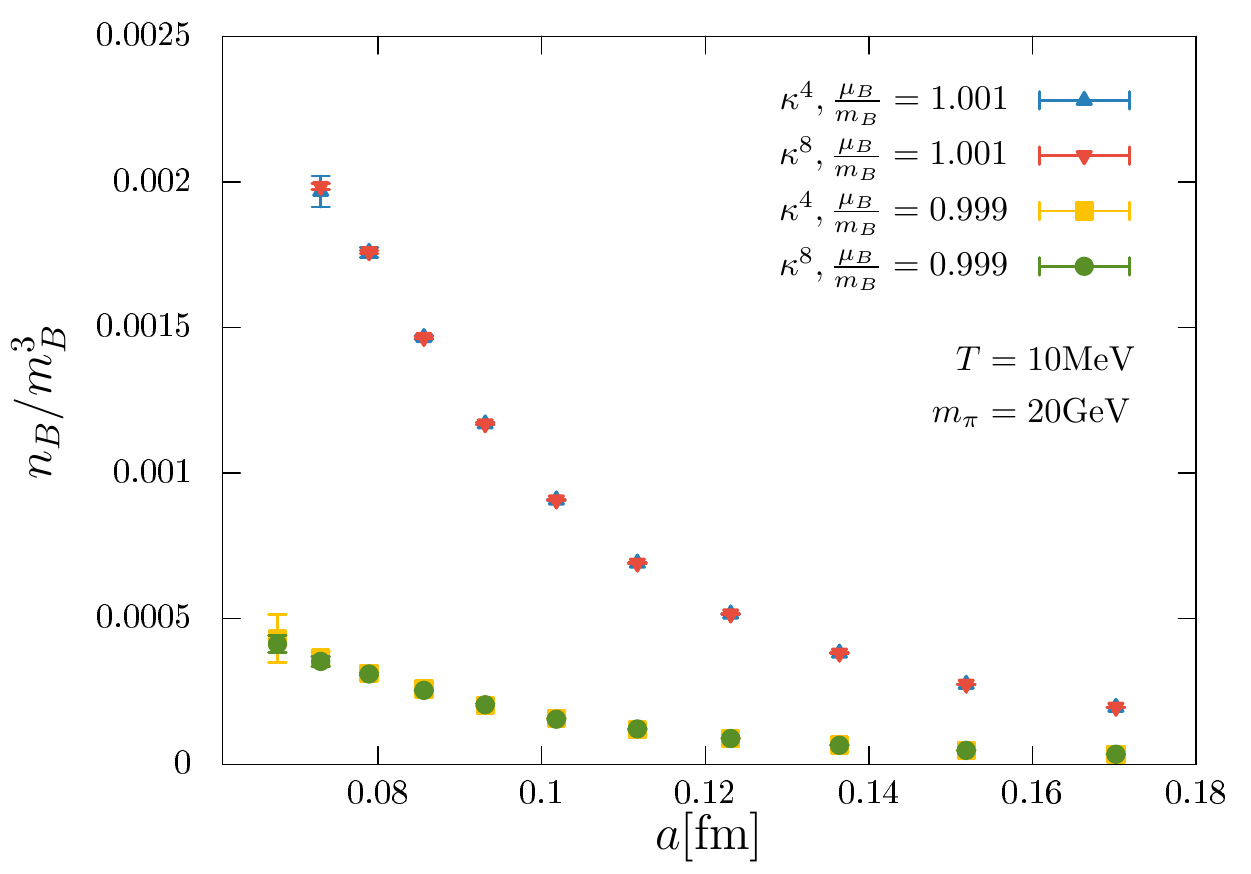}
\caption{Left: Baryon density in continuum units for various lattice spacings. Right: Continuum approach of the baryon density
for fixed chemical potentials. From \cite{k8}.}
\label{fig:cont}       % Give a unique label
\end{figure}

Lattice artefacts have to be removed by approaching the continuum. In figure \ref{fig:cont} (left)
the baryon density is shown in physical units for various lattice spacings. As expected, the lattice saturation level is pushed
higher for finer lattices and upon extrapolation will disappear to infinity. However, this implies rapidly exploding 
difficulties for any continuum extrapolation with growing density, as illustrated in figure \ref{fig:cont} (right) for two densities
near the onset transition. While the region of leading $O(a)$ lattice corrections for unimproved 
Wilson fermions appears to be reached, the curves become very steep with increasing chemical potential, making continuum 
extrapolations more difficult to control. Thus, enormously fine lattices are required before higher densities can be studied.
Note that this difficulty has nothing to do with the sign problem or the effective theory approach, but is a generic feature of
the discretisation. In other words, even if the sign problem could be overcome at no computational cost, densities higher than nuclear
would be out of reach for numerical simulations of full QCD.

\begin{figure}[t]
\centering
\includegraphics[width=4.5cm,clip]{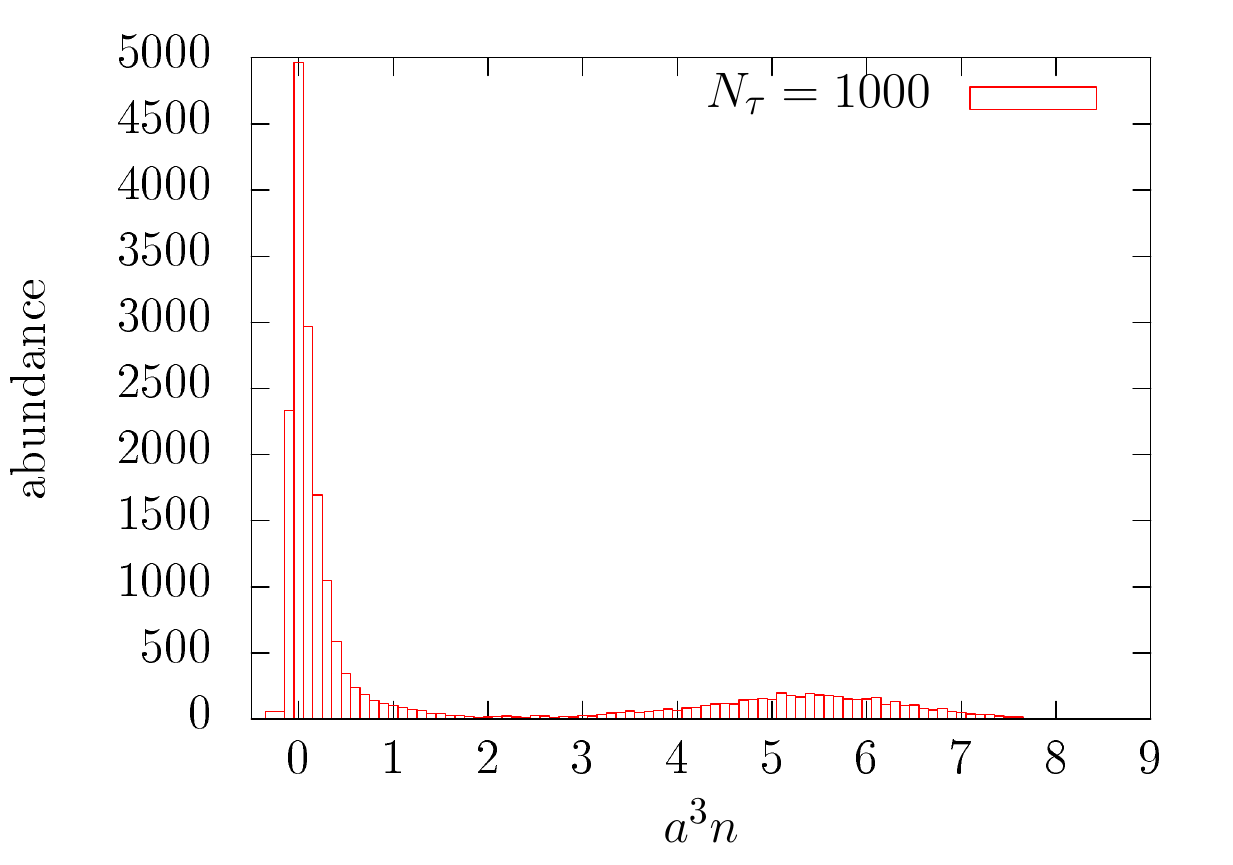}
\includegraphics[width=4.5cm,clip]{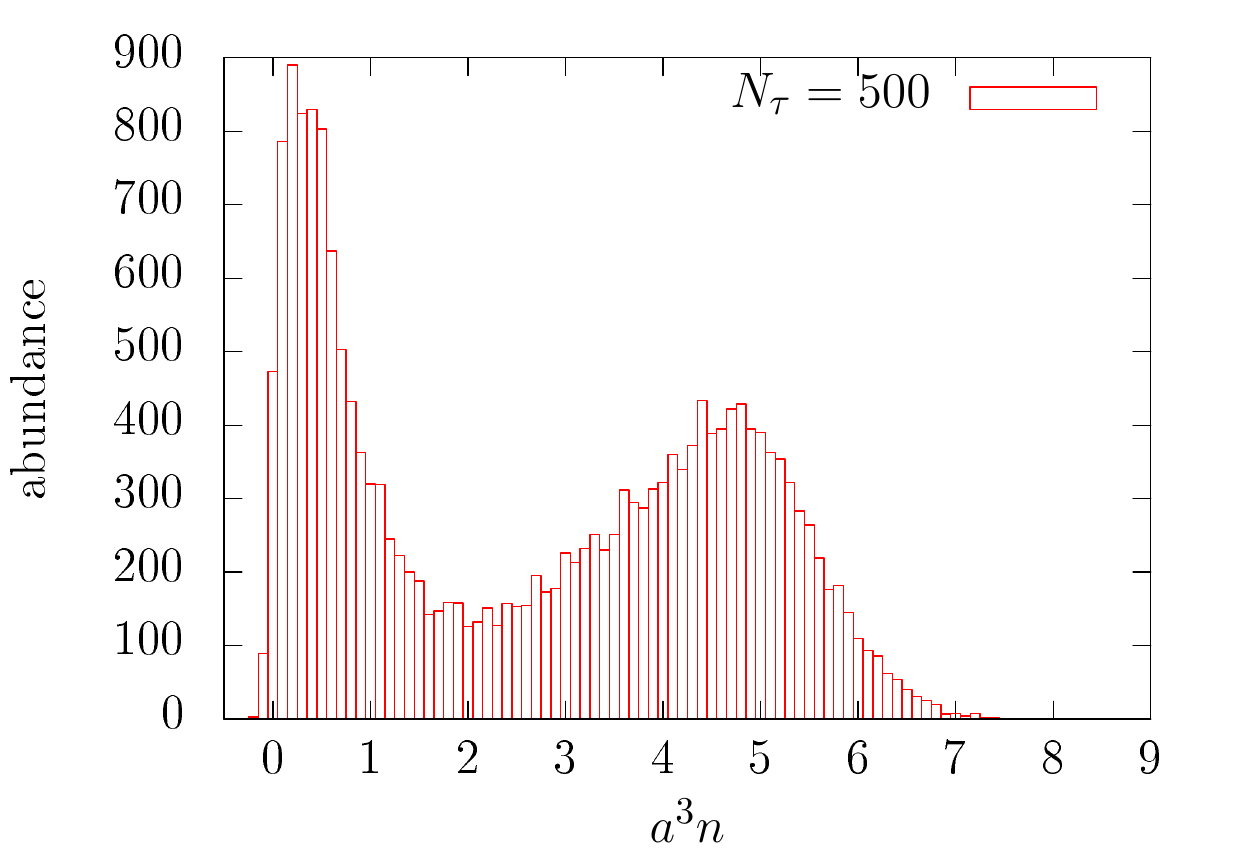}
\includegraphics[width=4.5cm,clip]{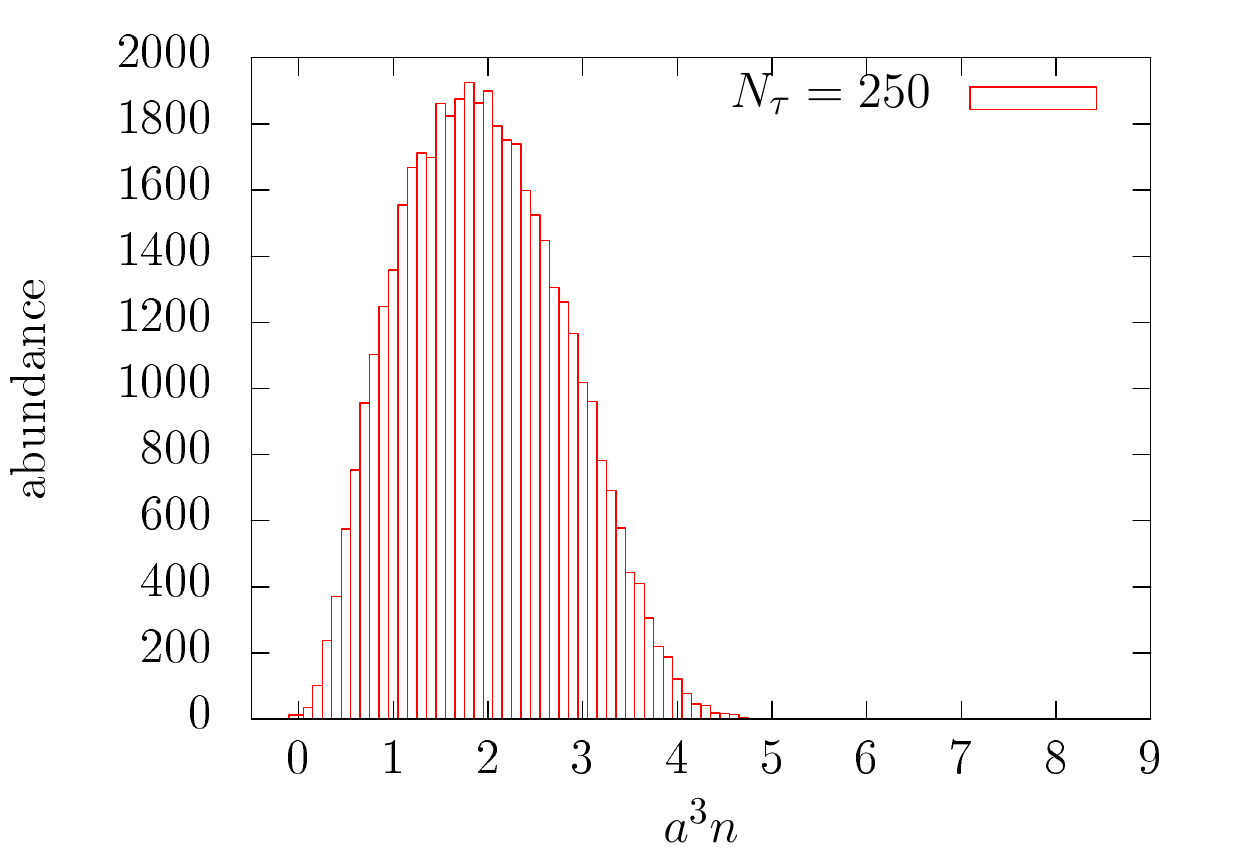}
\caption{Distributions of the quark density at the nuclear liquid gas transition for increasing temperature, $\kappa=0.12, \beta=5.7$.
The transition changes from first-order at low temperatures to crossover at higher temperatures. From \cite{bind}.}
\label{fig:nlg}       % Give a unique label
\end{figure}

Note that the onset
transition happens at $\mu_B^c\losim m_B$, due to the binding energy between the nucleons. Note
also that the onset transition here is a smooth crossover, in contrast to the first-order phase 
transition for physical QCD in nature. This is due to the fact that the binding energy per nucleon decreases
strongly as function of quark masses and for the heavy quarks studied here is smaller than 
the temperature realized in the previous figures. It was confirmed in \cite{bind} that a first-order
behavior indeed results for sufficiently small quark masses, as the histograms in figure \ref{fig:nlg} show
by changing from a two-peak structure at large $N_\tau$ (small temperature) to a Gaussian (crossover) 
with decreasing $N_\tau$ (increasing temperature).
However, in that mass range the effective
theory is not yet converging and terms of higher order in $\kappa$ are necessary to quantitatively
reproduce QCD. Nevertheless, the effective theory exhibits all qualitative features of the nuclear liquid gas
transition. 

\subsection{Analytic solution and equation of state}

\begin{figure}[t]
\centering
\includegraphics[width=5.5cm,angle=90,clip]{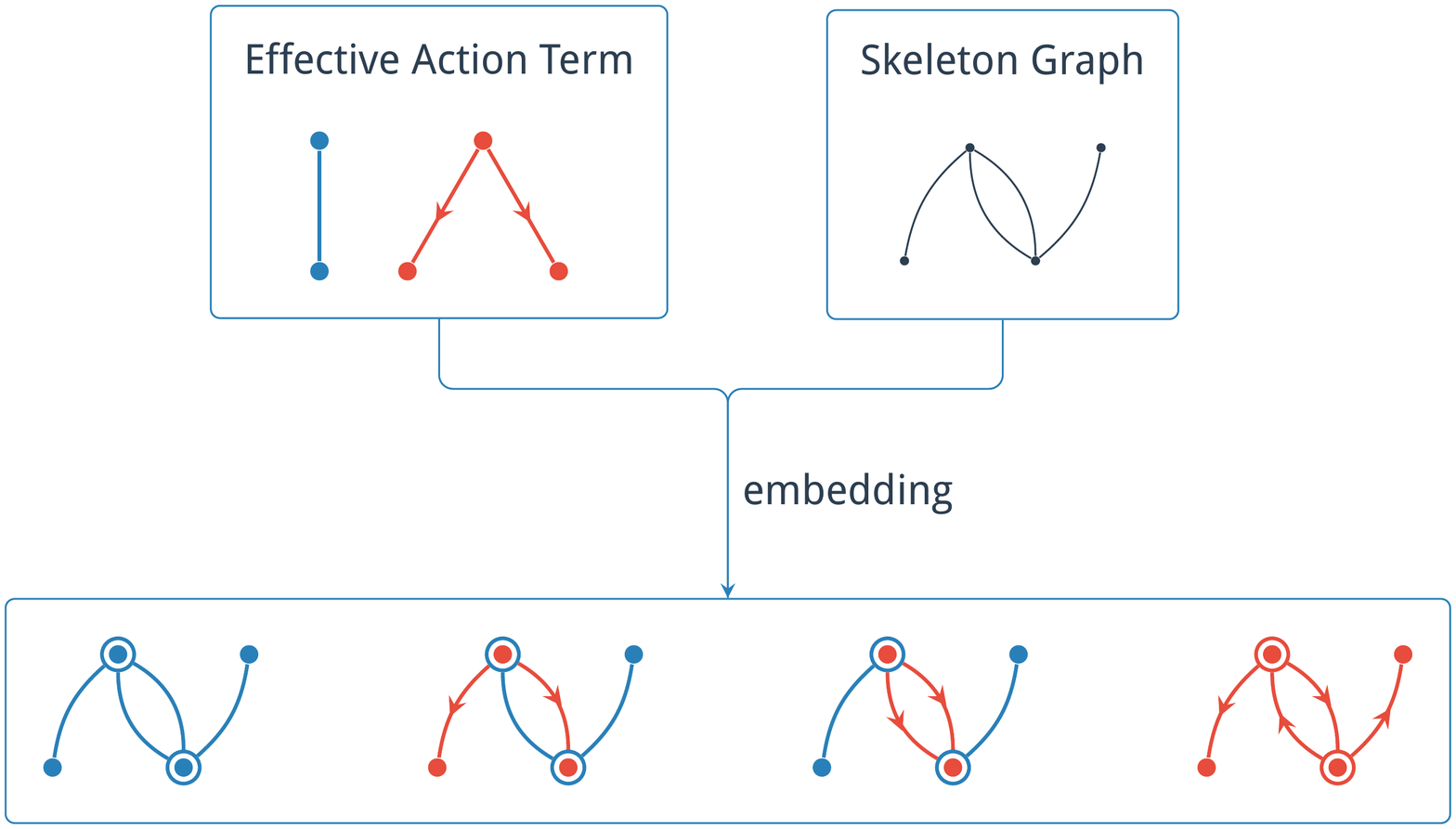}
\includegraphics[width=7cm,clip]{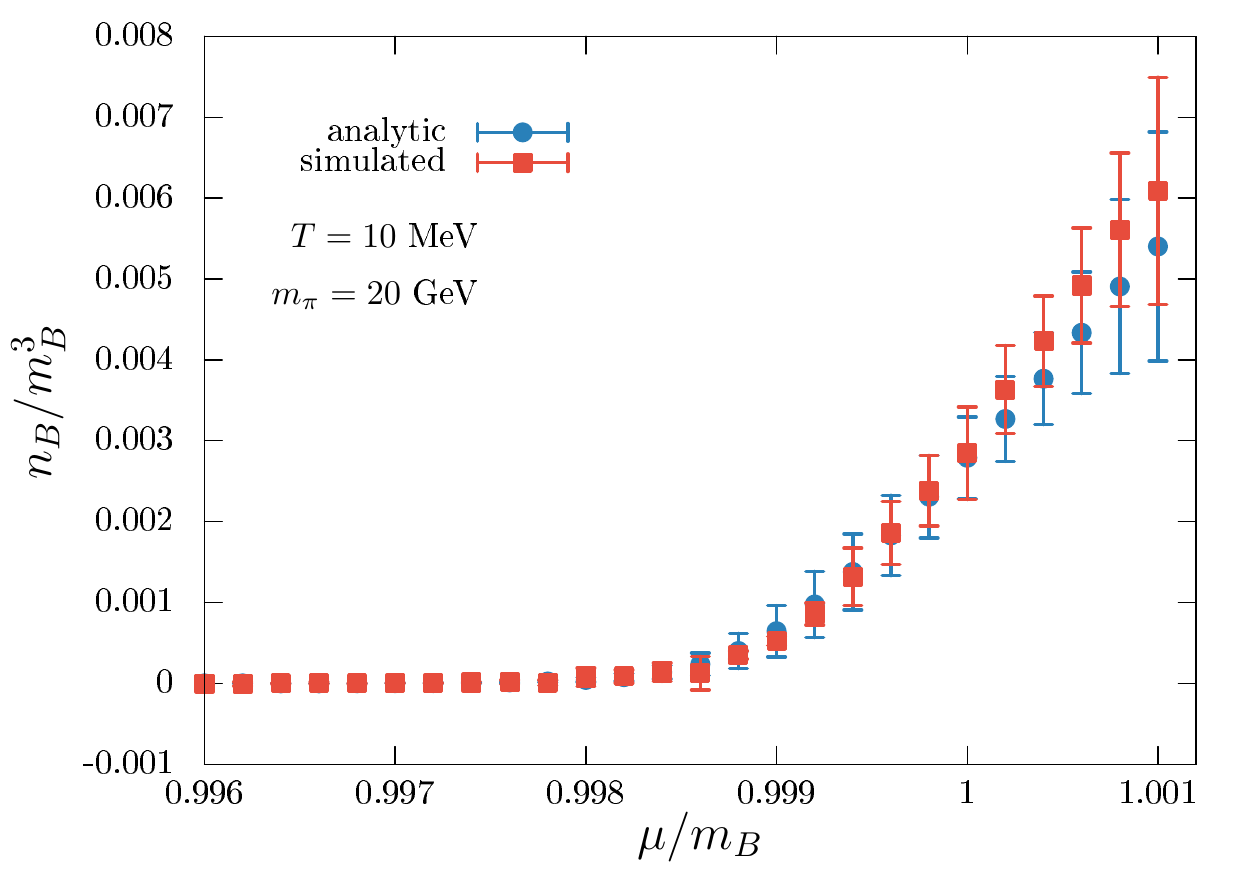}
\caption{Left: Mapping the non-local multi-point interactions of the effective lattice theory onto graphs of the linked cluster expansion.
Right: Continuum-extrapolated results for the $N_f=2$ baryon density from Langevin simulations and linked cluster expansion
of the 3d effective theory. From \cite{k8}.}
\label{fig:comp}       
\end{figure}

As first discussed in \cite{bergner,bind}, the effective couplings are small and correspond themselves to power series
in the original couplings $u(\beta),\kappa$, with the highest level of the effective theory in the cold and dense regime
being $u^5\kappa^8$ presently. Thus, in a range where the effective theory is valid, it should also permit
perturbative calculations in weak effective couplings. A systematic treatment can be achieved by noting that, since 
Polyakov loops are merely complex numbers,
the effective theory can be mapped into 
a spin model of the generalised form
\beq
  \mathcal{Z} = \int D\phi_i(x) \, \exp \bigg\{ -S_0[\phi_i] %
    - \frac{1}{2!} \sum_{x,y} v_{ij}(x,y) \phi_i(x) \phi_j(x)
  -\frac{1}{3!}\sum_{x,y,z} u_{ijk}(x,y,z) \phi_i(x) \phi_j(y) \phi_k(z) + \dots \bigg\}.
\eeq
It is then amenable to 
 linked cluster expansion techniques \cite{Wortis:1980}.
While the
standard linked cluster expansion is taylored to spin models with nearest neighbour 
interactions, here we have to deal with $n$-point interactions and larger distances as well. However,  
the geometry
of the terms in the effective action is exactly that of the set of connected graphs that can be embedded on a
square lattice. Employing a combinatorial embedding technique \cite{k8}, the calculation can be performed systematically
and fully analytically. Doing this for a sequence of lattice spacings and extrapolating to 
the continuum we compare with the numerical evaluation in Figure \ref{fig:comp} and find quantitative agreement. Hence
we now have a fully analytic way to compute thermodynamic functions of cold and dense QCD with heavy quarks, which does not 
suffer from the sign problem.

\begin{figure}[t]
\centering
\includegraphics[width=7cm,clip]{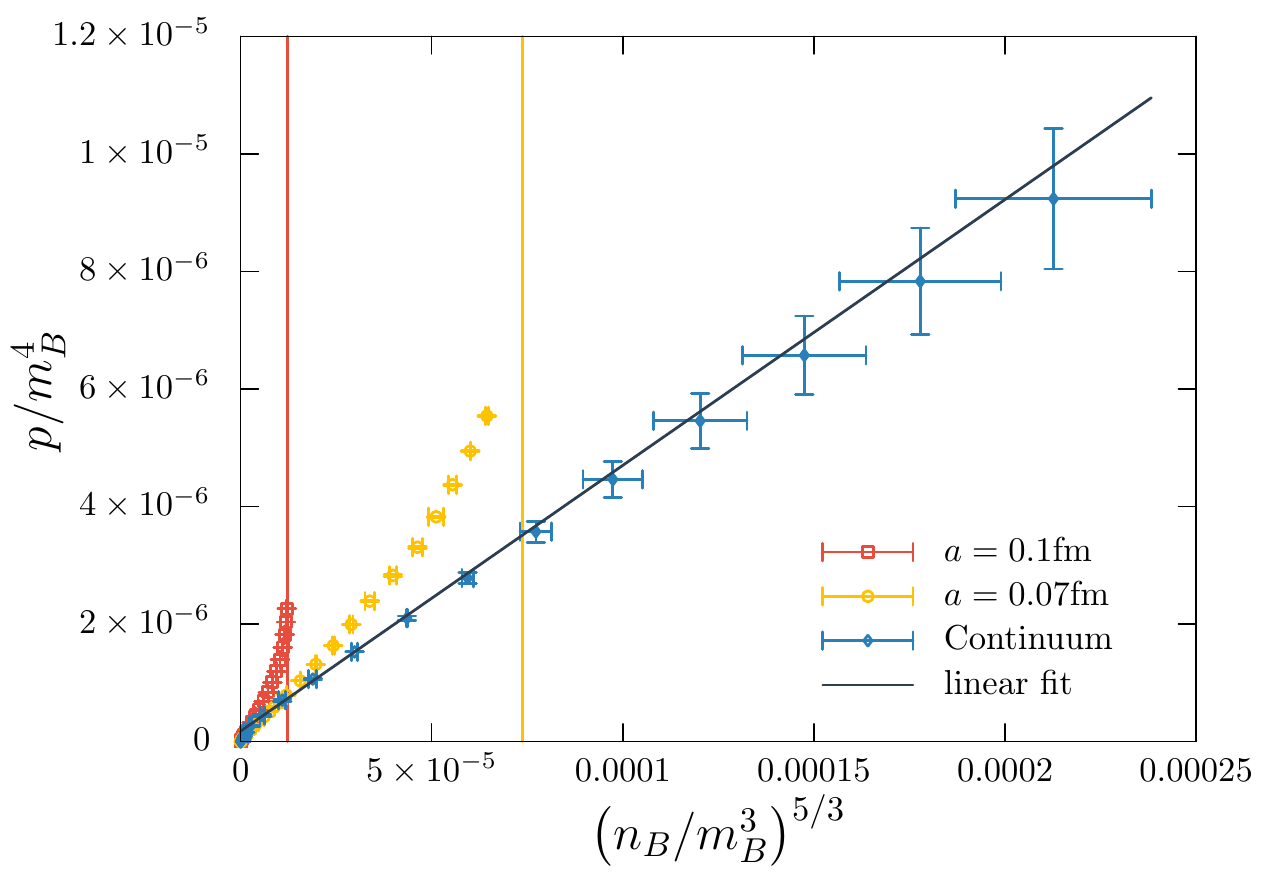}
\includegraphics[width=7cm,clip]{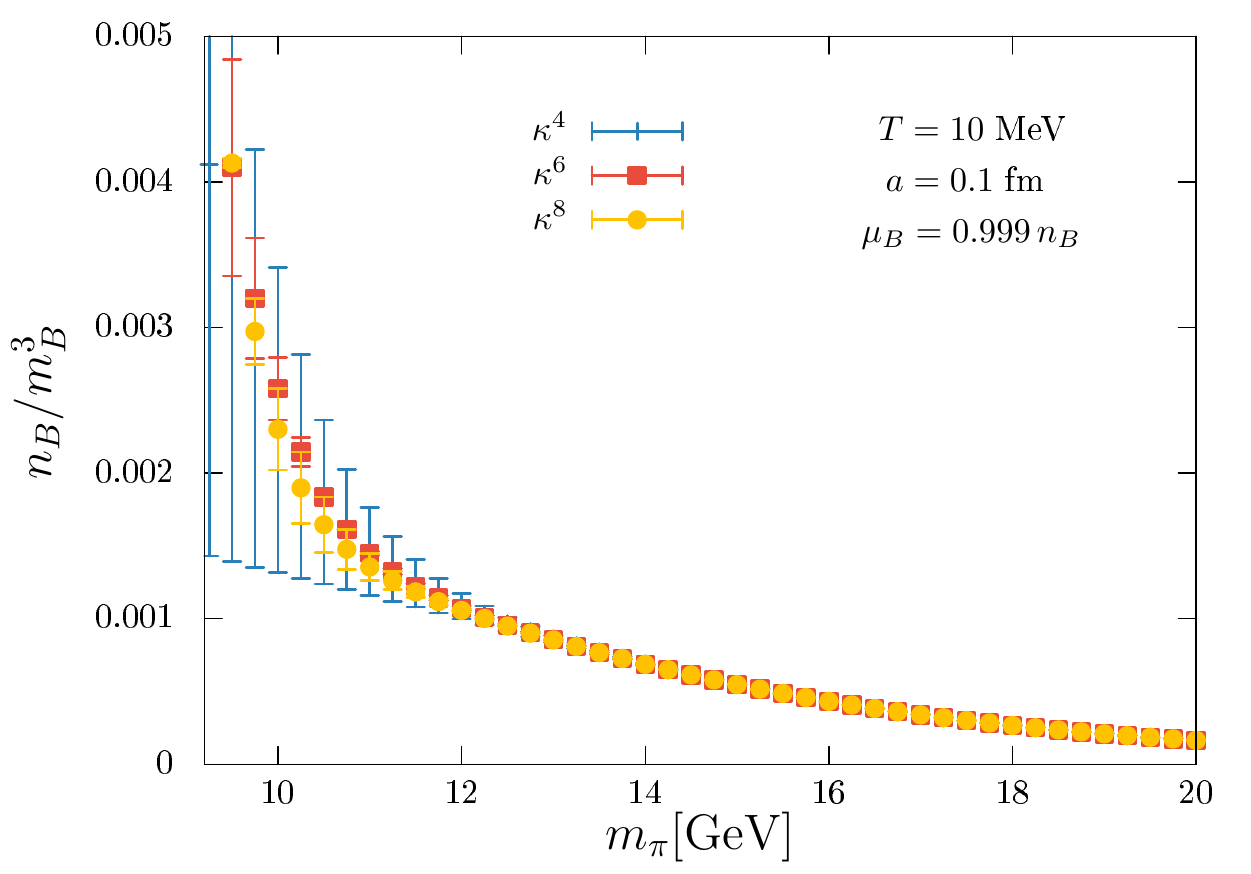}
\caption{Left: The continuum extrapolated pressure and baryon density are consistent with a polytropic equation of state 
for non-relativistic fermions. The vertical lines mark lattice saturation for fixed lattice spacings.
Right: Accessible mass range of the effective theory at a fixed lattice spacing. From \cite{k8}.}
\label{fig:eos}       % Give a unique label
\end{figure}

This calculational scheme can now be applied to compute the equation of state of baryonic matter.
Figure \ref{fig:eos} (left) shows the pressure as a function of baryon number density extrapolated to the continuum.
The straight line represents a fit to a polytropic equation of state corresponding to a free gas of non-relativistic 
fermions. From a physics point of view, this is not unexpected: the partition function contains a mix of bosonic and fermionic
baryons and we have already mentioned that the binding energy per baryon for heavy quarks is even smaller than in nature. For 
temperatures approaching zero the pressure is dominated by the fermionic contribution, hence the observed polytropic behaviour.
On the other hand, from a computational point of view this result is rather remarkable. We have started from quarks and gluons
in a strong coupling expansion, so the emergence of a weakly interacting baryon gas is a non-trivial dynamical feature.
Moreover, for any finite lattice spacing the equation of state is distorted by lattice saturation, marked by the vertical lines in 
figure \ref{fig:eos}, and not polytropic. The fact that the continuum extrapolated data sit on a physically sensible 
polytrope suggests that the different steps going into the extrapolation are under control.  

Finally one can investigate the accessible mass range of the effective theory by again assigning a systematic errorbar 
to the difference between results of the two highest orders in the hopping expansion. Figure \ref{fig:eos} (right) shows the
baryon density as a function of pion mass for a fixed lattice spacing and rapidly increasing error bars with decreasing mass.
There is an interplay between masses and lattice spacing: lighter masses are accessible on coarser lattices, but lattices have 
to be sufficiently fine if continuum results are desired.

\section{A 4d effective lattice theory for coarse lattices}

There is a second, and much older \cite{Wolff,Karsch}, effective lattice theory approach, which works in a  parameter region 
complementary to the previous one. Here the starting point is the staggered lattice formulation and the 
idea is to first integrate out all link variables to define an 
effective 4d theory in terms of fermionic fields only
\bea
Z&=&\int D\bar{\psi} D\psi DU_\mu \;e^{-S_g-S_f}\equiv \int D\bar{\psi} D\psi \;Z_F \langle e^{-S_g}\rangle_{Z_F},\qquad
Z_F=\int DU_\mu\; e^{-S_f}\;,\nn\\
\langle e^{-S_g}\rangle_{Z_F}&=&1-\langle S_g\rangle_{Z_F} + O(\beta^2)\;.
\eea
Again the defining first line is exact but not yet useful. In the second line the exponential of the gauge action, which is proportional
to the lattice coupling $\beta=2N_c/g^2$, is expanded in a strong coupling series.
In the strong coupling limit, $\beta=0$ or $g\rightarrow \infty$, this reduces to the first term and there is no contribution from the
pure gauge action.
After this step the quark fields form colour singlets, the mesons $M(x)=\bar{\psi}(x)\psi(x)$ 
and baryons
$B(x)=\frac{1}{6}\epsilon_{i_1 i_2 i_3}\psi_{i_1}(x)\psi_{i_2}(x)\psi_{i_3}(x)$.
In the second step, also the quarks are integrated out, which allows to express the partition function via integer variables:
\begin{align}
Z_{SC}(m_q,\mu)= \sum_{\{k_b,n_x,\ell\}}
\underbrace{\prod_{b=(x,\mu)}\frac{(N_c-k_b)!}{N_c!k_b!}}_{\text{meson hoppings}\, M_xM_y}
\underbrace{\prod_{x}\frac{N_c!}{n_x!}(2am_q)^{n_x}}_{
\text{chiral condensate}\, M_x}
\underbrace{\prod_\ell w(\ell,\mu)}_{\text{baryon hoppings}\, \bar{B}_xB_y}
\label{SCPF}
\end{align}
with $k_b\in \{0,\ldots 3\}$, $n_x \in \{0,\ldots 3\}$, $\ell_b \in \{0,\pm 1\}$. The Pauli principle for the quarks gets 
realised through a Grassmann constraint 
%\begin{equation}
%n_x+\sum_{\hnu=\pm\hat{0},\ldots, \pm \hat{d}}\lr{k_{\hnu}(x) + \frac{N_c}{2} |\ell_\hnu(x)|} = 3.
%\label{Grassmann}
%\end{equation}
which restricts the number of admissible configurations ${\{k_b,n_x,\ell\}}$ in (\ref{SCPF}) such that 
mesonic degrees of freedom always add up to $N_c$ and baryons form self-avoiding loops not in contact with
the mesons. The sign problem of this  effective theory can be completely cured by rearranging the partition function in an 
appropriate way. The theory is then amenable to very efficient simulation by worm algorithms, including the special
case of the chiral limit with vanishing quark mass, see e.g.~\cite{ff}.
\begin{figure}[t]
\centering
\includegraphics[width=6cm,clip]{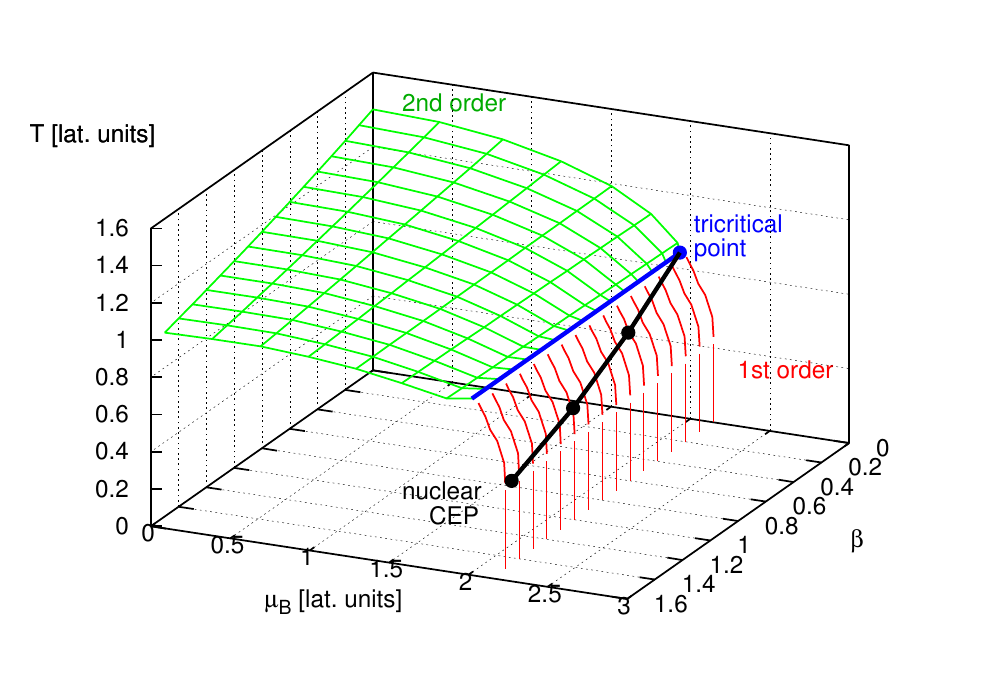}
\includegraphics[width=4cm,clip]{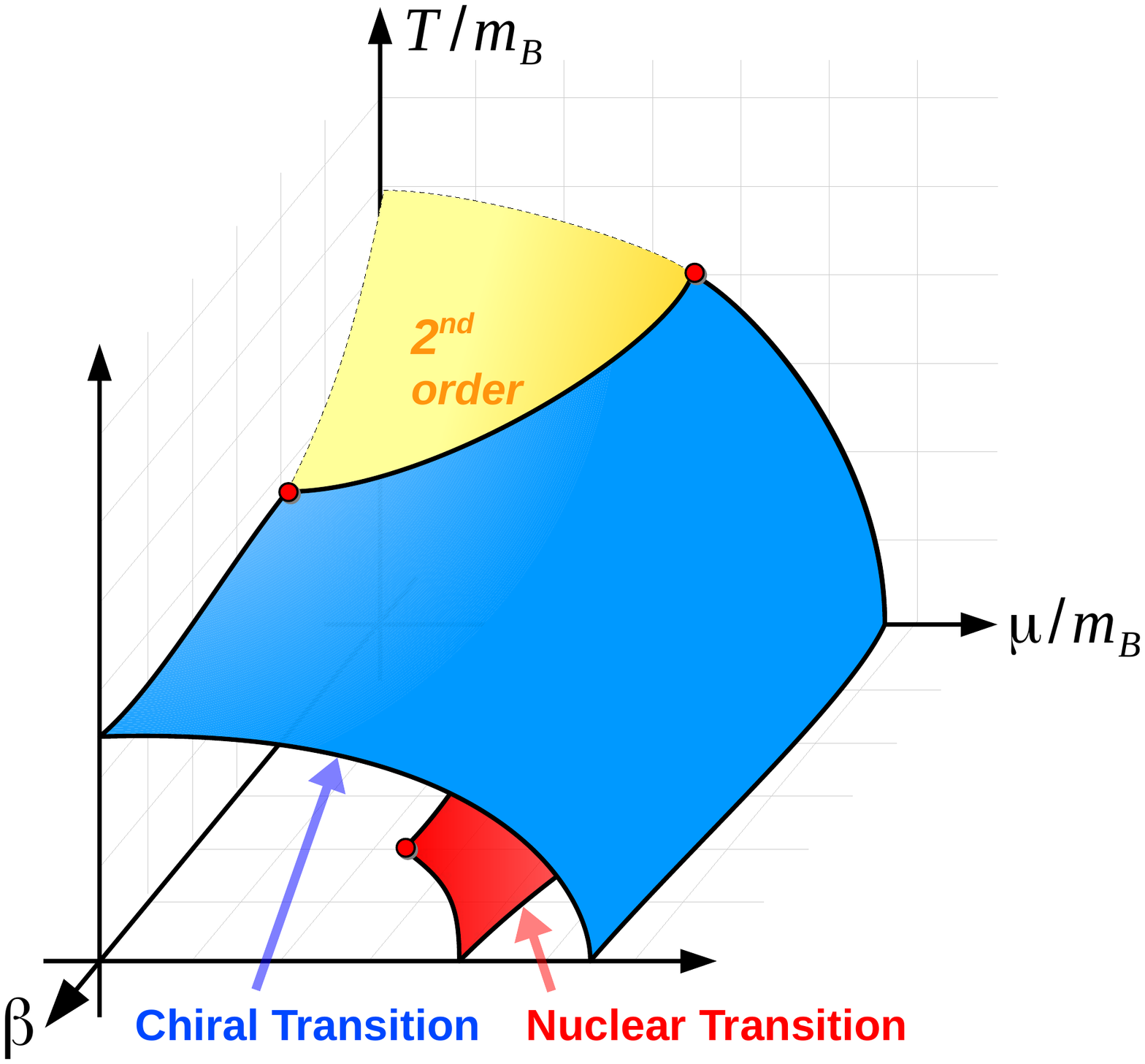}
\includegraphics[width=4cm,clip]{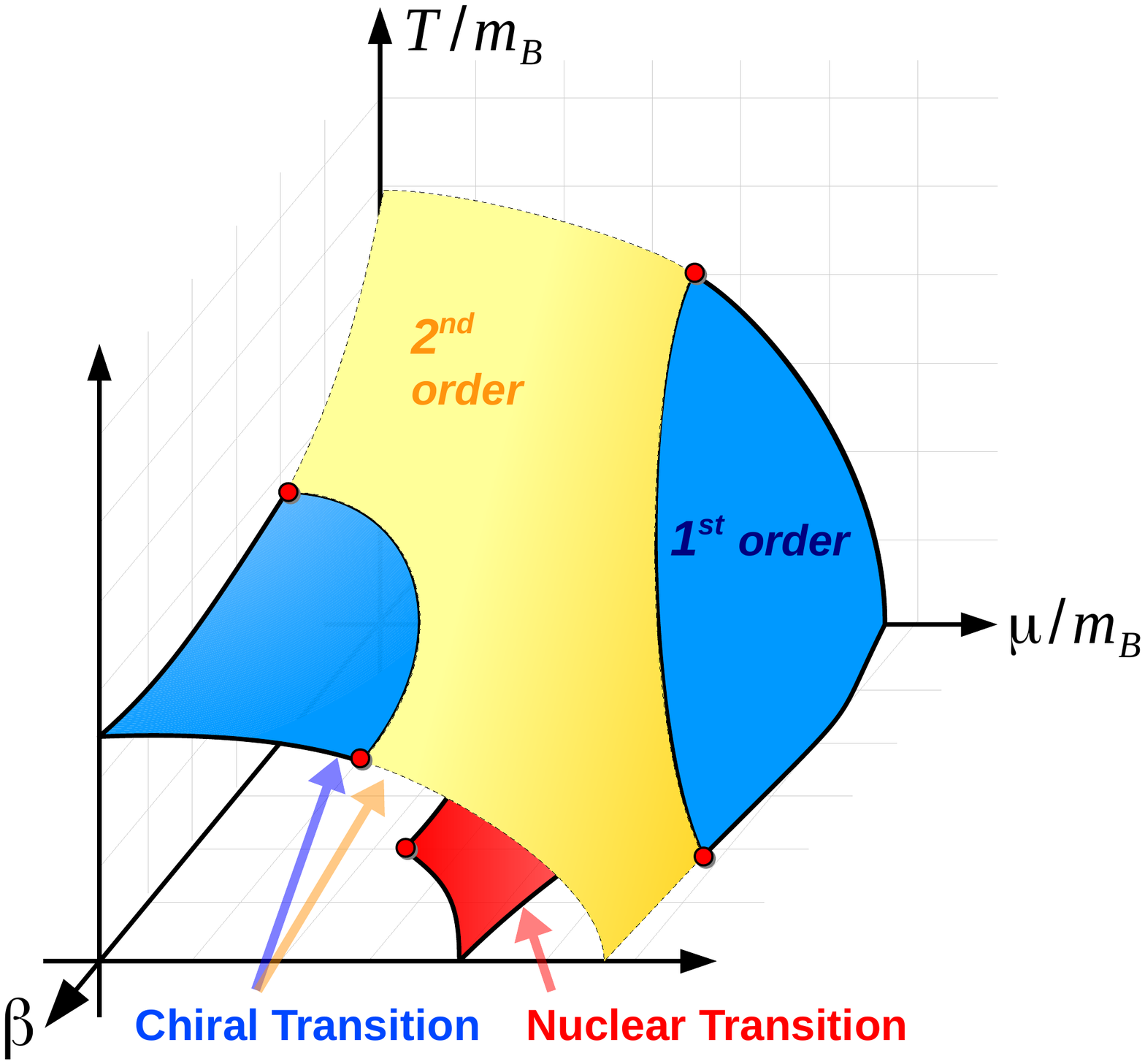}
\vspace*{-0.5cm}
\caption{Left: Phase diagram in lattice units for staggered quarks in the chiral limit and for strong coupling/coarse lattices.
Right: Two possible continuations to weak coupling/fine lattices. In the continuum limit $\beta=\infty$ the theory describes 
$N_f=4$ flavours. From \cite{wolfgang}.}
\label{fig:worm}      
\end{figure}

For zero mass quarks QCD has a global chiral symmetry, and the staggered lattice formulation preserves a subgroup of this.
As a result, a true non-analytic phase transition separates the phases of broken and restored chiral symmetry, which
is second order for small chemical potentials and turns first order beyond a tricritical point, as illustrated in the 
$\beta=0$ backplane of figure \ref{fig:worm} (left). Note that the onset transition to nuclear matter in the strong coupling limit
coincides with the chiral transition, with its endpoint sitting on top of the chiral tricritical point. 
More recently the leading gauge corrections to $O(\beta)$ have been evaluated as well \cite{unger}. At finite values of $\beta$
the two transitions begin to become distinct. Figure \ref{fig:worm} (right) shows two of the various possible continuations to
 larger $\beta$-values. Note that in the continuum limit $\beta\rightarrow \infty$ the theory describes $N_f=4$ flavours because
 of the lattice doublers in the staggered formulation, hence the transition at $\mu=0$ is constrained to be of first order.

\section{Conclusions}

Strong coupling methods permit to analytically compute effective lattice theories in which the sign problem can be 
dealt with and which allow a description of the cold and dense regime of QCD.  These theories are valid in complementary
parameter regimes for heavy quarks close to the continuum and for light/chiral quarks on coarse lattices, respectively.
In particular, both approaches give a description of baryon condensation and nuclear matter directly from QCD. This 
is encouraging to further pursue the challenging task of extending the effective theories towards the physical regime. \\

\noindent
{\bf Acknowledgements:}
This presentation is based on various collaborations with G.~Bergner, Ph.~de Forcrand, 
M.~Fromm, J.~Glesaaen, J.~Langelage, M.~Neuman, S.~Lottini and W.~Unger,
and supported by the Helmholtz International Center for FAIR within the 
LOEWE program of the State of Hesse.

\end{document}